# Study on the Vibration Displacement Distribution of a Circular Ultrasonic Motor Stator

Chol-Su Ri[a)], Myong-Jin Kim[a)], Chol-Su Kim[b)], Song-Jin Im[b)]

[a)] Chair of Acoustics, Department of Physics, Kim Il Sung University, Pyongyang, DPR Korea

[b)] Chair of Optics, Department of Physics, Kim Il Sung University, Pyongyang, DPR Korea

**Abstract**

In this paper is presented a theoretical consideration on the stator's displacement distribution, which is one of the most important problems in defining the structure of the circular ultrasonic motor stator. The results are compared with results obtained utilizing holographic interferometer, laser vibrometer and a FEM (finite element method) simulation. They are in a good agreement with each other.

## 1. Introduction

Ultrasonic motors(USMs) are attracting considerable attentions due to a number of their potential applications. Ultrasonic motors, compared to conventional electromagnetic motors, have excellent performances and many superior features such as high holding torque, high torque at low speed, quiet operation, simple structure, compactness in size, high resolution of displacement control(good controllability), low consume, stability in terrible environment and no electromagnetic radiation and noise. For this reason, USM are widely used as actuators for robots, rover, auto-focus camera lenses, precise positioning devices, machines in space and medical treatment devices, etc.[1-5]

Many kinds of ultrasonic motor have been proposed in the past. And they are now in the stages of practical application as an alternative to electromagnetic motors.

Generally, there are several geometrical types of ultrasonic actuators: ring, cylinder, disk, etc.[6-8,14]. Many studies related to ultrasonic motors were reported, but most of them were included in the USMs with ring type. The theoretical analysis on USMs can be performed by using mathematical models[9,10] or finite element method[11] to improve and modify the mechanical efficiency.[12]

USM uses two energy conversions to generate output torque: First, electrical energy from power source leads up to the elliptic motion on the stator surface by piezoelectric effect and secondly, the microscopic vibration of the stator is converted to macroscopic motion of the rotor by friction between the stator and the rotor.

In this paper a theoretical model of a circular USM stator, which actually is a disk with a hole in its center, is proposed. The results are compared with those measured by time average holographic interferometry and laser vibrometer, and simulated by FEM(finite element method).

## 2. Theoretical Modelling of a circular USM stator

Figure 1 shows the shape of a plastic circular plate with a hole in center. Here, the origin of the coordinate system coincides with the center of the hole.

In general, vibration equation of a thin plate is as follow:[13]

$$\nabla^4 \xi(r,\theta,t) + \frac{12\rho(1-v^2)}{Eh^2} \cdot \frac{\partial^2 \xi(r,\theta,t)}{\partial t^2} = 0, \qquad (1)$$

where $\rho$, $v$, $h$, $E$ are the density, the Poisson's ratio, the thickness and the Young's modulus of the plate respectively.

Solution of Eq. (1) can be written as

$$\xi(r,\theta,t) = \eta(r,\theta) \cdot e^{j\omega t}. \qquad (2)$$

Substituting Eq. (2) into Eq. (1) results in

$$\nabla^4 \eta(r,\theta) - \beta^4 \eta(r,\theta) = 0, \qquad (3)$$

where

$$\beta^4 = \frac{12\rho(1-v^2)\omega^2}{Eh^2} \qquad (4)$$

The general solution of Eq. (3) is equal to the linear combination of solutions of the following two equations:

$$\begin{cases} (\nabla^2 + \beta^2)\eta_1(r,\theta) = 0 \\ (\nabla^2 - \beta^2)\eta_2(r,\theta) = 0 \end{cases} \qquad (5)$$

$$\eta(r,\theta) = \eta_1(r,\theta) + \eta_2(r,\theta) \qquad (6)$$

The two solutions $\eta_1(r,\theta)$ and $\eta_2(r,\theta)$ of Eq.(5), are

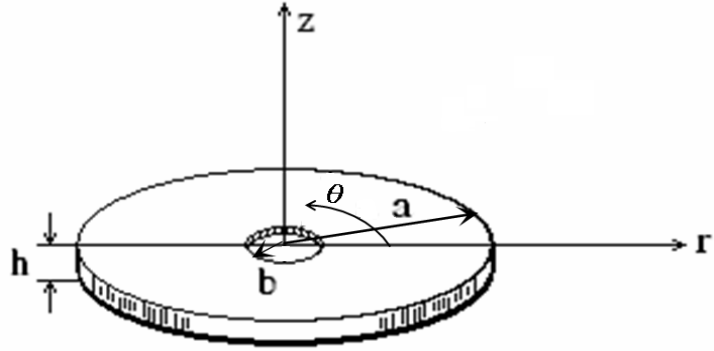

Fig. 1. Elastic circular place with a hole in center.

$$\begin{cases} \eta_1(r,\theta) = [A_n J_n(\beta,r) + B_n N_n(\beta,r)]\cos n\theta \\ \eta_2(r,\theta) = [C_n I_n(\beta,r) + D_n K_n(\beta,r)]\cos n\theta \end{cases} \quad (7)$$

Then, the general solution of Eq. (1) is

$$\xi(r,\theta,t) = [A_n J_n(\beta r) + B_n N_n(\beta r) + \\ + C_n I_n(\beta r) + D_n K_n(\beta r)]\cos n\theta \cdot e^{j\omega t}, \quad (8)$$

where $J_n$, $N_n$ are the nth Bessel function and Neumann function respectively and $I_n$, $K_n$ the nth imaginary Bessel function and imaginary Neumann function respectively. $A_n, B_n, C_n, D_n$ are coefficients which must be determined from the boundary condition.

In general, two conditions, bending momentum $M$ and shear force $Q$, are $M|_r = 0$ and $Q|_r = 0$ at free ends. The inner and outer boundaries are free as shown in Fig. 1, so the boundary condition here is

$$\begin{cases} M|_r = -D\left[\dfrac{\partial^2 \eta}{\partial r^2} + v\left(\dfrac{\partial \eta}{\partial r} + \dfrac{1}{r^2}\dfrac{\partial^2 \eta}{\partial \theta^2}\right)\right]_{r=a,b} = 0 \\ Q|_r = N|_{r=a,b} + \dfrac{1}{r}\dfrac{\partial H_r}{\partial \theta}\bigg|_{r=a,b} = 0 \end{cases}, \quad (9)$$

where

$$\begin{cases} H_r = -(1-\nu)\dfrac{\partial}{\partial r}\left(\dfrac{1}{r}\dfrac{\partial \xi}{\partial}\right)\cdot D \\ N_r = -D\dfrac{\partial \Delta \xi}{\partial r} \end{cases}, \qquad (10)$$

and $D$ is the stiffness of plate material.

Coupling expression (8) and boundary condition (9) results in a matrix equation concerned with coefficients $A_n, B_n, C_n, D_n$:

$$[H_{ij}]\begin{bmatrix} A_n \\ B_n \\ C_n \\ D_n \end{bmatrix} = 0, \qquad (11)$$

where

$$[H_{ij}] = \begin{bmatrix} j_1 & n_1 & i_1 & k_1 \\ j_2 & n_2 & i_2 & k_2 \\ j_3 & n_3 & i_3 & k_3 \\ j_4 & n_4 & i_4 & k_4 \end{bmatrix},$$

$j_1 = J_n(\beta a) - (1-\nu)[n(n-1)J_n(\beta a)/(\beta a)^2 + J_{n+1}(\beta a)/\beta a],$

$n_1 = N_n(\beta a) - (1-\nu)[n(n-1)N_n(\beta a)/(\beta a)^2 + N_{n+1}(\beta a)/\beta a],$

$i_1 = -I_n(\beta a) - (1-\nu)[n(n-1)I_n(\beta a)/(\beta a)^2 - I_{n+1}(\beta a)/\beta a],$

$k_1 = -K_n(\beta a) - (1-\nu)[n(n-1)K_n(\beta a)/(\beta a)^2 + K_{n+1}(\beta a)/\beta a]$

$j_2 = J_n(\beta b) - (1-\nu)[n(n-1)J_n(\beta b)/(\beta b)^2 + J_{n+1}(\beta b)/\beta b],$

$n_2 = N_n(\beta b) - (1-\nu)[n(n-1)N_n(\beta b)/(\beta b)^2 + N_{n+1}(\beta b)/\beta b],$

$i_2 = -I_n(\beta b) - (1-\nu)[n(n-1)I_n(\beta b)/(\beta b)^2 - I_{n+1}(\beta b)/\beta b],$

$k_2 = -K_n(\beta b) - (1-\nu)[n(n-1)K_n(\beta b)/(\beta b)^2 + K_{n+1}(\beta b)/\beta b]$

$j_3 = nJ_n(\beta a) - \beta a J_{n+1}(\beta a) + n^2(1-\nu)[(n-1)J_n(\beta a) - \beta a J_{n+1}(\beta a)]/(\beta a)^2,$

$n_3 = nN_n(\beta a) - \beta a N_{n+1}(\beta a) + n^2(1-\nu)[(n-1)N_n(\beta a) - \beta a N_{n+1}(\beta a)]/(\beta a)^2,$

$i_3 = -nI_n(\beta a) - \beta a I_{n+1}(\beta a) + n^2(1-\nu)[(n-1)I_n(\beta a) + \beta a I_{n+1}(\beta a)]/(\beta a)^2,$

$k_3 = nK_n(\beta a) - \beta a K_{n+1}(\beta a) + n^2(1-\nu)[(n-1)K_n(\beta a) - \beta a K_{n+1}(\beta a)]/(\beta a)^2$

$j_4 = nJ_n(\beta b) - \beta b J_{n+1}(\beta b) + n^2(1-\nu)[(n-1)J_n(\beta b) - \beta b J_{n+1}(\beta b)]/(\beta b)^2,$

$n_4 = nN_n(\beta b) - \beta b N_{n+1}(\beta b) + n^2(1-\nu)[(n-1)N_n(\beta b) - \beta b N_{n+1}(\beta b)]/(\beta b)^2,$

$i_4 = -nI_n(\beta b) - \beta b I_{n+1}(\beta b) + n^2(1-\nu)[(n-1)I_n(\beta b) + \beta b I_{n+1}(\beta b)]/(\beta b)^2,$

$$k_4 = nK_n(\beta b) - \beta b K_{n+1}(\beta b) + n^2(1-\nu)[(n-1)K_n(\beta b) - \beta b K_{n+1}(\beta b)]/(\beta b)^2$$

In order that Eq. (11) has a solution of which all components aren't zero simultaneously, it must be satisfied

$$|H_{ij}| = 0 \qquad . \qquad (12)$$

Setting the solutions of Eq. (12) $\beta_{nm}$ ($n, m = 0, 1, 2, \ldots$), and $A_n = 1$, $B_n$, $C_n$ and $D_n$ are

$$B_n = \frac{\begin{vmatrix} -j_2 & i_2 & k_2 \\ -j_3 & i_3 & k_3 \\ -j_4 & i_4 & k_4 \end{vmatrix}}{\begin{vmatrix} n_2 & i_2 & k_2 \\ n_3 & i_3 & k_3 \\ n_4 & i_4 & k_4 \end{vmatrix}}, \quad C_n = \frac{\varepsilon\begin{vmatrix} n_2 & -j_2 & k_2 \\ n_3 & -j_3 & k_3 \\ n_4 & -j_4 & k_4 \end{vmatrix}}{\begin{vmatrix} n_2 & i_2 & k_2 \\ n_3 & i_3 & k_3 \\ n_4 & i_4 & k_4 \end{vmatrix}}, \quad D_n = \frac{\varepsilon\begin{vmatrix} n_2 & i_2 & -j_2 \\ n_3 & i_3 & -j_3 \\ n_4 & i_4 & -j_4 \end{vmatrix}}{\begin{vmatrix} n_2 & i_2 & k_2 \\ n_3 & i_3 & k_3 \\ n_4 & i_4 & k_4 \end{vmatrix}}. \qquad (13)$$

Therefore, the solution of Eq. (1) with boundary conditions is

$$\xi(r,\theta,t) = [A_n J_n(\beta_{nm}r) + B_n N_n(\beta_{nm}r) + C_n I_n(\beta_{nm}r) + D_n K_n(\beta_{nm}r)]\cos n\theta \cdot e^{j\omega t} \qquad (14)$$

The vibration displacement distributions of the stator calculated numerically by using Eq. (14) is shown in Fig. 2. Here outer radius is $a = 21.5mm$ and inner radius $b = 2mm$.

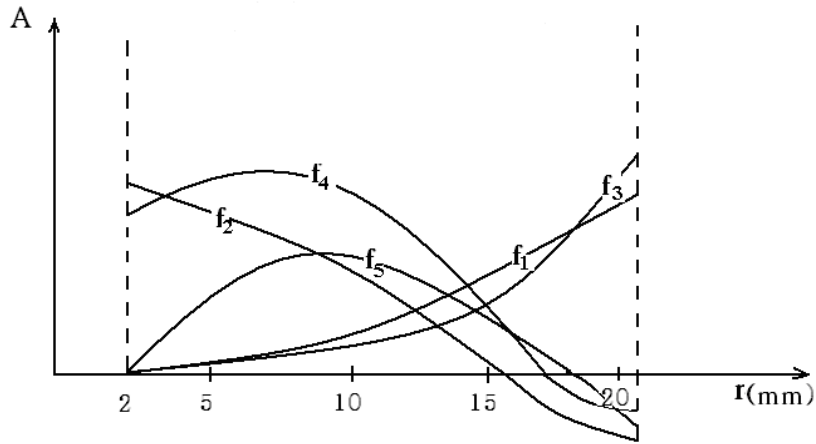

Fig. 2. Radial distribution of vibration distribution of a stator ($\theta = 0$), a) $f_1 = 7.7kHz$, b) $f_2 = 12.7kHz$, c) $f_3 = 17.4kHz$, d) $f_4 = 30.5kHz$, e) $f_5 = 49.5kHz$

## 3. Experimental Measurement and Simulation

Fig. 3 shows block diagram for measurement of vibration displacement by time-average holographic interferometry.

In general, intensity distribution on the reconstructed image of a vibrating plate obtained with holographic interferometry is proportional to the square of the 0-th Bessel function of first kind.[15]

$$I \sim J_0^2(\vec{k} \cdot \vec{A}) \tag{15}$$

where $\vec{K} = \vec{K}_2 - \vec{K}_1$ is sensitivity vector, $\vec{K}_1$ and $\vec{K}_2$ are wave number vectors in the direction of observation and illumination respectively and $\vec{A}$ is vibration displacement vector of the object surface.

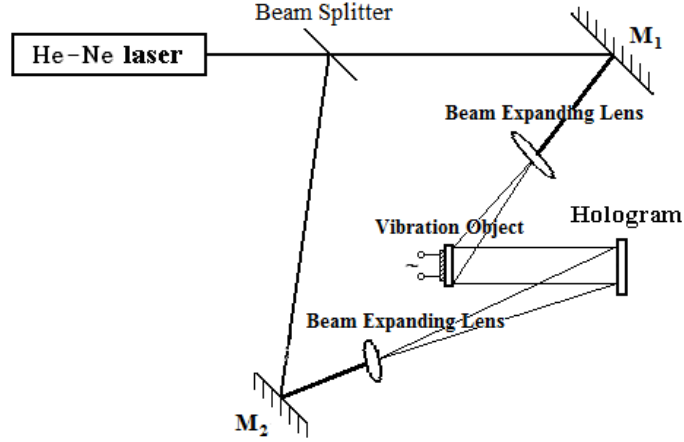

Fig. 3. Measurement of vibration displacement of USM stator by holographic interferometer.

According to expression (15), changing of directions of illumination light and observation result in movement of interference fringe pattern. The vibration displacement on the vibration surface can be determined by analyzing the interferogram formed on object surface. A scanning laser vibrometer(Polytec OFV 3001S) was also employed to measure the vibration displacement distribution of a circular USM stator in the condition of free ends for five frequencies, $f_1 \sim f_5$.

Finite element analysis is performed by using commercial package (ANSYS). The finite element method is very useful in finding the resonance frequency and analyzing the vibrational displacement distribution of vibration objects with any geometrical shapes and dimensions, so the USM stator can be modeled and simulated by ANSYS.

The material of the circular plate of the stator is hard aluminum. The segmented ceramic elements which is and attached on the circular plate for excitation is an equivalent of PZT-4, and the standard material parameters of PZT element are used. The material parameters and geometrical dimensions of the circular plate are as follows: $E = 7.02 \times 10^{10} N/m^2$ ,

$\rho = 2.70 \times 10^3 kg/m^3$, $\nu = 0.34$, $C = 5.10 \times 10^3 m/s$, $a = 21.5mm$, $b = 2mm$, $h = 2.5mm$.

Material parameters and dimensions of PZT element are $\rho_0 = 7.6 \times 10^3 kg/m^3$, $c_0 = 2.95 \times 10^3 m/s$, $a_0 = 21.5mm$, $b_0 = 2mm$, $h_0 = 0.5mm$. Harmonic response analysis is performed in order to find the vibration displacement distribution on the surface of the circular USM stator. An ideal voltage generator which provides 1V voltage is used to drive the PZT element. Internal structural losses of the transducer have been taken into account by applying an appropriate damping ratio(0.15%), which is defined as the ratio of the loss energy over the kinetic energy when harmonic response analysis is done. The analysis frequencies are $7.7kHz, 12.7kHz, 17.4kHz, 30.5kHz$ and $49.5kHz$.

Fig. 4 and Fig. 5 show the experimental interference pattern and the simulation results, respectively. As shown in two figures, holographic reproduction image is very similar to simulation result.

The vibration modes at different frequencies measured by holographic interferometry and simulated with ANSYS FEM coincide with each other completely.

Vibration displacement distributions of stator $(\theta = 0)$, measured by the holography interferometer and a laser vibrometer(Polytec OFV 3001S), are in a good agreement, and they have similar forms to theoretical curves as in Fig. 2. Positions of minima and maxima of vibration displacements at 5 frequencies are shown in Table.1.

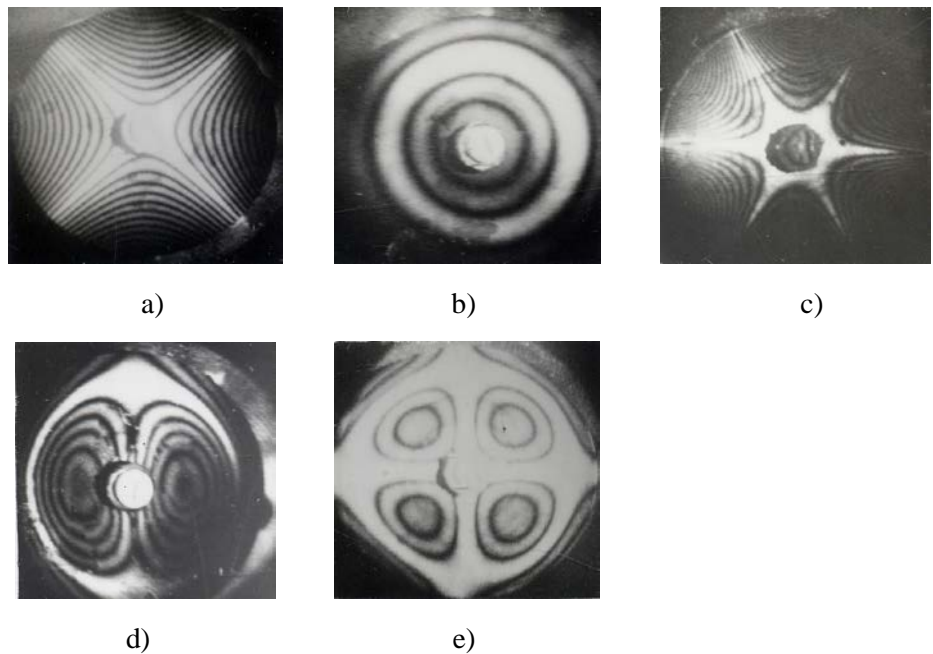

Fig. 4. Holographic reproduction image of a circular ultrasonic motor stator. a) $f_1 = 7.7kHz$, b) $f_2 = 12.7kHz$, c) $f_3 = 17.4kHz$, d) $f_4 = 30.5kHz$, e) $f_5 = 49.5kHz$

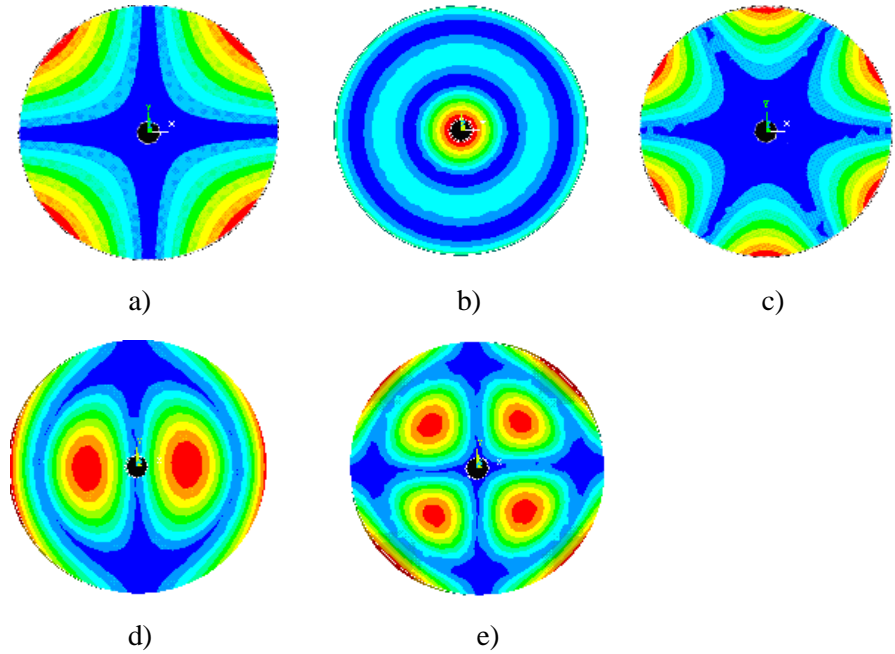

Fig. 5. Distribution of vibration displacement of a circular ultrasonic motor stator(ANSYS FEM),
a) $f_1 = 7.7 kHz$, b) $f_2 = 12.7 kHz$, c) $f_3 = 17.4 kHz$, d) $f_4 = 30.5 kHz$, e) $f_5 = 49.5 kHz$

Table.1. Comparison between theoretical and experimental results

of minima and maxima of vibration displacement.

| F(kHz) | 7.7 | | 12.7 | | 17.4 | | 30.5 | | 49.5 | |
|---|---|---|---|---|---|---|---|---|---|---|
| | $r_{max}$ | $r_{max}$ | $r_{min}$ | $r_{max}$ | $r_{min}$ | $r_{max}$ | $r_{min}$ | $r_{max}$ | $r_{min}$ | $r_{max}$ |
| Theory | 2 | 21.5 | 15 | 2 | 2 | 21.5 | 16.4 | 8.4 | 17.9 | 10.3 |
| Experiment | 2 | 21.5 | 15.1 | 2 | 2 | 21.5 | 16.5 | 8.5 | 18.0 | 10.5 |

## Conclusions

In this paper has been considered a theoretical method for determining the distribution of bending vibration displacement on the circular USM stator with a hole in center. Vibration displacements of a circular USM stator have been measured by time-average holographic interferometer and laser vibrometer, and simulated with ANSYS FEM. The results of measurements and simulation were compared with the theoretical result. The results are in a good agreement with each other.